\documentclass[twocolumn]{aastex6}
\slugcomment{Submitted to AJ, October 2016, revised xxx 201x}
\shorttitle{Extinction in the Cluster SAI~113}
\shortauthors{Carraro et al.}

\begin{document}

\title{Extinction in the Star Cluster SAI~113 and Galactic Structure in Carina}

\author{Giovanni Carraro}
\affil{Dipartimento di Fisica e Astronomia, Universit\'{a} di Padova, Vicolo Osservatorio 3, I-35122, Padova, Italy}
\email{giovanni.carraro@unipd.it}

\author{David~G. Turner and Daniel~J. Majaess}
\affil{Department of Astronomy and Physics, Saint Mary's University, Halifax, NS B3H 3C3, Canada}
\email{turner@ap.smu.ca}
\email{dmajaess@ns.sympatico.ca}

\and

\author{Gustavo L. Baume, Roberto Gamen, and Jos\'{e} A. Molina Lera}
\affil{Facultad de Ciencias Astron\'{o}micas y Geof\'{i}sicas (UNLP), IALP (CONICET, UNLP), Paseo del Bosque s/n, La Plata, Argentina}
\email{gbaume@fcaglp.unlp.edu.ar}
\email{rgamen@fcaglp.unlp.edu.ar}
\email{jalejom1@fcaglp.unlp.edu.ar}

\begin{abstract}
Photometric CCD {\it UBVI}$_C$ photometry obtained for 4860 stars surrounding the embedded southern cluster SAI~113 (Skiff~8) is used to examine the reddening in the field and derive the distance to the cluster and nearby van~Genderen~1. Spectroscopic color excesses for bright cluster stars, photometric reddenings for A3 dwarfs, and dereddening of cluster stars imply that the reddening and extinction laws match results derived for other young clusters in Carina: E$_{U-B}/$E$_{B-V} \simeq 0.64$ and $R_V \simeq 4$. SAI~113 displays features that may be linked to a history of dynamical interactions among member stars: possible circumstellar reddening and rapid rotation of late B-type members, ringlike features in star density, and a compact core with most stars distributed randomly across the field. The group van~Genderen~1 resembles a stellar asterism, with potential members distributed randomly across the field. Distances of $3.90 \pm0.19$ kpc and $2.49 \pm0.09$ kpc are derived for SAI~113 and van~Genderen~1, respectively, with variable reddenings E$_{B-V}$ ranging from 0.84 to 1.29 and 0.23 to 1.28. The SRC variables CK~Car and EV~Car may be outlying members of van~Genderen~1, thereby of use for calibrating the period-luminosity relation for pulsating M supergiants. More importantly, the anomalous reddening and extinction evident in Carina and nearby regions of the Galactic plane in the fourth quadrant impact the mapping of spiral structure from young open clusters. The distribution of spiral arms in the fourth quadrant may be significantly different from how it is often portrayed. 
\end{abstract}


\section{Introduction}
Researchers at the Sternberg Astronomical Institute (SAI) in Moscow recently compiled initial parameters for a sample of about 170 newly-detected and previously-unstudied open star clusters \citep{gl10} discovered from analyses of 2MASS {\it JHK}$_s$ photometry \citep{cu03} for Galactic fields covered by the 2MASS survey \citep{sk06}, many subjected to additional study using {\it UBVI} data. Object 113 in the sample, SAI~113, at ($\alpha_{2000}$, $\delta_{2000}$) = (10:22:43.6, --59:30:20), is one of 23 such groups separated into a table of ``embedded clusters,'' for which parameters are not given, typically because of large scatter in the 2MASS colors. SAI~113 is the primary subject of the present study.

The cluster was noticed previously by Brian Skiff while deriving co-ordinates for stars near the luminous blue variable HR Carinae observed by \citet{vg91}. The brightest stars in SAI~113 are numbered 78, 82, 83, 84, and 85 on an inset of van~Genderen's Fig.~1, which is reversed from the field's appearance in the sky. The group is therefore also designated as Skiff~8 in the most recent version of Bruno Alessi's open cluster catalogue \citep{al16}.

A separate, sparse group of early-type stars is visible a few arcminutes east of HR Car on the \citet{vg91} chart at ($\alpha_{2000}$, $\delta_{2000}$) = (10:22:29, --59:38:30), and is designated as van~Genderen~1 \citep{al16}. SAI~113 and van~Genderen~1 are both fairly compact, and members of both display differential reddening according to the transformed Walraven photometry for the stars tabulated by \citet{vg91}, as well as by the dispersion in their 2MASS colors. The present study was initiated in order to fill the void in our knowledge of both groups, with an initial focus on SAI~113, particularly given the location of the two groups in the rich complex of young stars, open clusters, and associations belonging to and surrounding the Eta Carinae complex.

\begin{figure}[h]
\epsscale{1.08}
\plotone{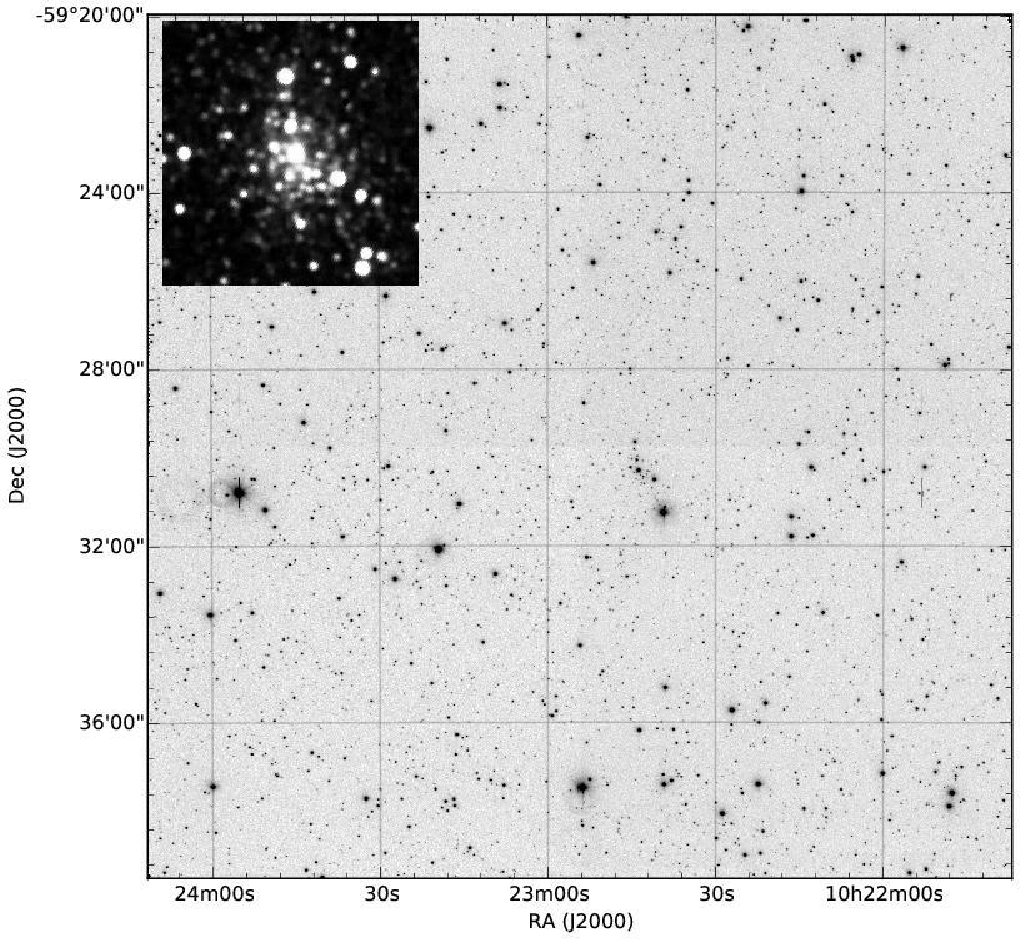}
\caption{The $20^{\prime} \times 20^{\prime}$ field of SAI~113 centered on ($\alpha_{2000}$, $\delta_{2000}$) = (10:22:54, --59:30:00), as viewed in blue light, with north up and east to the left. The clump of stars at the cluster core, shown separately in the inset according to its appearance in 2MASS colorized images, lies immediately southwest of the field center, while van~Genderen~1 lies just above the lower right margin of the field.}
\label{fig1}
\end{figure}

A blue light image centered roughly on SAI~113 is presented in Fig.~\ref{fig1}, where it can be noted that the term ``embedded'' refers to the appearance on colorized images of the 2MASS survey \citep{sk06}, displayed as a BW inset to Fig.~\ref{fig1}, of the small clump of faint stars at its core, immediately lower right of image center. van~Genderen~1 lies above the lower right margin of Fig.~\ref{fig1}. Except for the latter, bright stars display no obvious concentration towards either group. Rather, they are more-or-less evenly distributed across the field, possibly concentrating towards the margins of Fig.~\ref{fig1}, akin to the ringlike distribution in the cluster Collinder~70, the ellipsoidal group of stars surrounding $\epsilon$ Orionis and referred to as the Orion Stellar Ring \citep[see][]{sc71}, as well as to the ringlike concentration of stars surrounding the anonymous cluster near HD~18391 \citep{te09}. Both seem to be in advanced stages of dissolution into the surrounding field, which may also be true of SAI~113 and van~Genderen~1.

\begin{figure}[b]
\epsscale{1.00}
\plotone{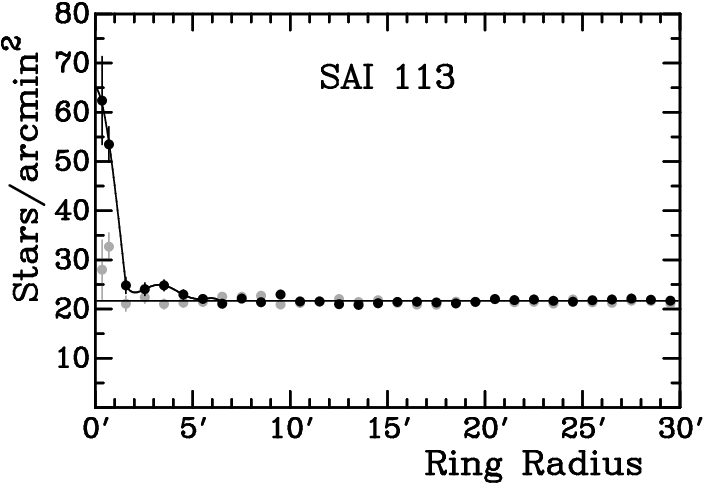}
\caption{Star counts off the 2MASS survey to 30 arcminutes from the centers for SAI~113 (black points) and van~Genderen~1 (gray points), with uncertainties tied to Poisson statistics. The field level was established from counts in the 15--30 arcminute rings. The curve is a schematic illustrating likely variations in star density for the inner regions of SAI~113.}
\label{fig2}
\end{figure}

Star counts were made for SAI~113 to the limits of the 2MASS survey, as illustrated in Fig.~\ref{fig2}. Stars were counted in rings about the cluster center of \citet{gl10}, following the statistical approach of \citet{vb60}. One arcminute rings were adopted, except for an additional half arcminute ring at the cluster center. According to the counts, the cluster has a nuclear radius of $r_n = 1^{\prime}.5$ and a coronal radius of $R_c = 6^{\prime}$ in the notation of \citet{kh69}, with $244\pm51$ cluster members lying above the field level of $21.68\pm0.05$ stars armin$^{-2}$ within $6^{\prime}$ of the cluster center. The subtle blips in the counts around $3^{\prime}$ and $9^{\prime}$ from the cluster center may be real, since they lie several $\sigma$ above the field star level. They suggest the existence of inner ringlike structures such as those in the outer regions of Collinder~70 and Anon~(HD~18391), other dissolving clusters. Otherwise, outlying stars tend to be distributed randomly across the field, more like an association than an open cluster.

Similar star counts were made for van~Genderen~1 using the cluster center cited previously, with results plotted in Fig.~\ref{fig2} as gray points. van~Genderen~1 is clearly a less prominent group than SAI~113, and may represent merely the dissolved remains of a former young cluster. That is also the impression obtained from casual examination of the field on Sky Survey images. The counts imply the existence of just $4\pm17$ group members above field level within 2$^{\prime}$ of the adopted center.

\begin{figure}[h!]
\epsscale{1.00}
\plotone{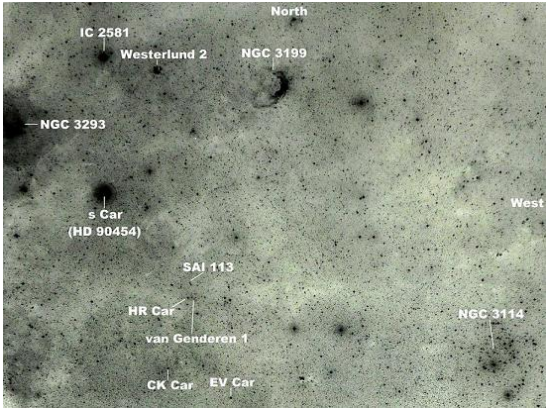}
\caption{The roughly $3\degr.5 \times 5\degr$ field west of the Eta Carinae complex containing SAI~113 and other objects discussed in the text. Note the numerous opaque dust clouds running across the field that seem to merge near SAI~113. The image is from a $J$-band (blue) exposure from the ESO-SRC survey.}
\label{fig3}
\end{figure}

The location of SAI~113 and van~Genderen~1 in the Carina complex is of particular interest because of the possibility of using them to test the extent of the ``anomalous'' extinction that has been documented for most of the clusters and stellar groups in the region \citep{tu12a,ca13}. According to \citet{nk80}, the extinction for this direction of the Galaxy arises in dust clouds roughly 1 kpc distant, while deep images of the field in blue light (Fig.~\ref{fig3}) reveal that strands of the same dust cloud permeate all of the ``anomalous'' regions of Carina, including those that converge near SAI~113. Prominent dust strands in Fig.~\ref{fig3} can be traced northwards to Westerlund~2 and IC~2581, and eastward to NGC~3293 and (outside the image) to the clusters of the Great Carina Nebula: Trumpler~14, Trumpler~15, Trumpler~16, Collinder~228, Collinder~236, Ruprecht~93, and Anon~(WR~38/38a) \citep{tu12a}. The anomalies relate to a shallow reddening slope of E$_{U-B}$/E$_{B-V} \simeq 0.64$ on the Johnson system and a large ratio of total-to-selective extinction of $R_V = A_V$/E$_{B-V} \simeq 4$. The anomaly for $R_V$ is particularly important since it affects the derived distances to clusters in Carina: 1.9--2.3 kpc relative to original estimates of 2.5--3.4 kpc \citep[see][]{tu12a}. Such a change dramatically affects how we picture the width and orientation of the Carina-Sagittarius arm of the Galaxy.

\section{Observations and Data Reduction}

{\it UBVI}$_C$ images centered on SAI~113 were obtained on two nights ($17\,\&\,22$) of March 2009 at Cerro Tololo Inter-American Observatory (CTIO) using the Y4KCAM camera on the 1.0-m telescope operated by the SMARTS consortium, as noted in Table~\ref{tab1}. The camera was equipped with an STA~4064 $\times$ 4064 CCD with $15 \mu$ pixels, a set-up that provided direct imaging over a field of view (FOV) of $20^{\prime}.0 \times 20^{\prime}.0$ at a scale of 0$^{\prime\prime}.289$ pixel$^{-1}$. The typical full-width half-maximum (FWHM) for the data was about $0^{\prime\prime}.9$. Fig.~\ref{fig1} is based on a 60-second exposure of the field in the {\it B} band.

\begin{deluxetable}{@{\extracolsep{+2.5mm}}lcll}
\tabletypesize{\small}
\tablewidth{0pt}
\centering
\tablenum{1}
\tablecaption{{\it UBVI} Photometric Imaging of SA~113 in March 2009
\label{tab1}}
\tablehead{
\colhead{Date} &\colhead{Filter} &\colhead{Exposures (s)} &\colhead{Airmass X} }
\startdata
March 17 &{\it U} &30, 200, 2000 &1.17--1.18 \\
&{\it B} &20, 150, 1500 &1.21--1.22 \\
&{\it V} &10, 100, 900 &1.32--1.34 \\
&{\it I} &10, 100, 900 &1.27--1.28 \\
\noalign{\smallskip}
March 22 &{\it U} &30, 200, 2000 & 1.16 \\
&{\it B} &20, 150, 1500 &1.19 \\
&{\it V} &10, 100, 900 &1.26--1.27 \\
&{\it I} &10, 100, 900 &1.23--1.24 \\
\enddata
\end{deluxetable}

The CTIO images were processed with the IRAF packages CCDRED and DAOPHOT using the point spread function (PSF) method of \citet{st87} to extract instrumental magnitudes. The PSF for each image was obtained for several isolated, spatially well-distributed, bright stars in the field ($\sim20$ in the sample). The PSF photometry for individual objects was then aperture corrected for each filter, where the corrections were computed from aperture photometry on the same stars used as PSF models. All resulting tables were combined by the DAOMASTER code \citep{st92}, where only those objects with $\chi^{2} \leq 3$ and abs(sharp) $\leq 0.5$ were included in order to avoid spurious detections. The functions $\chi^{2}$ and abs(sharp) are standard diagnostics in DAOPHOT used in automatic flux detection mode to measure the quality of a PSF fit relative to the adopted PSF model, and to monitor the brightness of detections and avoid defects caused by bad pixels. The night 22/03/09 was used as a photometric reference.

Our {\it UBVI}$_C$ data were calibrated through observations of about 70 standard stars in \citet{la92} fields SA~98 and PG~1047+003, the former providing very good color coverage. Aperture photometry was then carried out for all stars using the PHOTCAL package with transformation equations of the form:
\begin{eqnarray}
u = U + u1 + u2 (U-B) + u3 X                  \\
b = B + b1 + b2 (B-V) + b3 X                  \\
v = V + v1_{bv} + v2_{bv} (B-V) + v3_{bv} X   \\
v = V + v1_{vi} + v2_{vi} (V-I) + v3_{vi} X   \\
i = I_C + i1 + i2 (V-I) + i3 X                
\end{eqnarray}
\noindent
where {\it UBVI}$_C$ and {\it ubvi} are standard and instrumental magnitudes, respectively, {\it X} is the airmass of the observation, and typical values for the extinction coefficients for CTIO were adopted \citep[see][]{ba11}. Equation (3) was used to derive {\it V} magnitudes when the {\it B} magnitude was available, otherwise equation (4) was used. The calibration coefficients for the night 22/03/2009 are summarized in Table~\ref{tab2}. 

\begin{deluxetable}{@{\extracolsep{+0.5mm}}ccccrl}
\tabletypesize{\small}
\tablewidth{0pt}
\tablenum{2}
\tablecaption{Calibration Co-efficients for March 22, 2009 \label{tab2}}
\tablehead{\colhead{} &\colhead{} &\colhead{} &\colhead{} &\colhead{} &\colhead{}}
\startdata
$u1$ &$-0.851 \pm 0.006$ &$b1$ &$-2.064 \pm 0.010$ \\
$u2$ &$0.45$ &$b2$ &$0.25$ \\
$u3$ &$-0.029 \pm 0.009$ &$b3$ &$0.128 \pm 0.009$ \\
rms &$\pm 0.08$ &rms &$\pm 0.08$ \\
\noalign{\smallskip}\noalign{\smallskip}
$v1_{bv}$ &$-2.129 \pm 0.008$ &$v1_{vi}$ &$-2.119 \pm 0.006$ \\
$v2_{bv}$ &$0.16$ &$v2_{vi}$ &$0.16$ \\
$v3_{bv}$ &$-0.021 \pm 0.008$ &$v3_{vi}$ &$-0.032 \pm 0.005$ \\
rms  &$\pm 0.06$ &rms &$\pm 0.02$ \\
\noalign{\smallskip}\noalign{\smallskip}
$i1$ &$-1.321 \pm 0.005$  \\
$i2$ &$0.08$  \\
$i3$ &$-0.016 \pm 0.004$  \\
rms &$\pm 0.04$  \\
\enddata
\end{deluxetable}

World Co-ordinate System (WCS) header information was obtained for each image using the ALADIN tool and 2MASS data \citep{sk06}. \citet{ba09} describe the procedure used to perform an astrometric calibration of the data. That yielded reliable astrometry with uncertainties in position of order $\sim0^{\prime\prime}.12$.

The STILTS tool was used to manipulate tables and to cross-correlate the {\it UBVI}$_C$ and  2MASS {\it JHK}$_s$ data. The result was a catalogue with astrometric/photometric information on the detected objects covering a FOV of approximately $20^{\prime} \times 20^{\prime}$. The complete catalogue will be made available in electronic form at the CDS website.

\begin{figure}[h!]
\epsscale{1.00}
\plotone{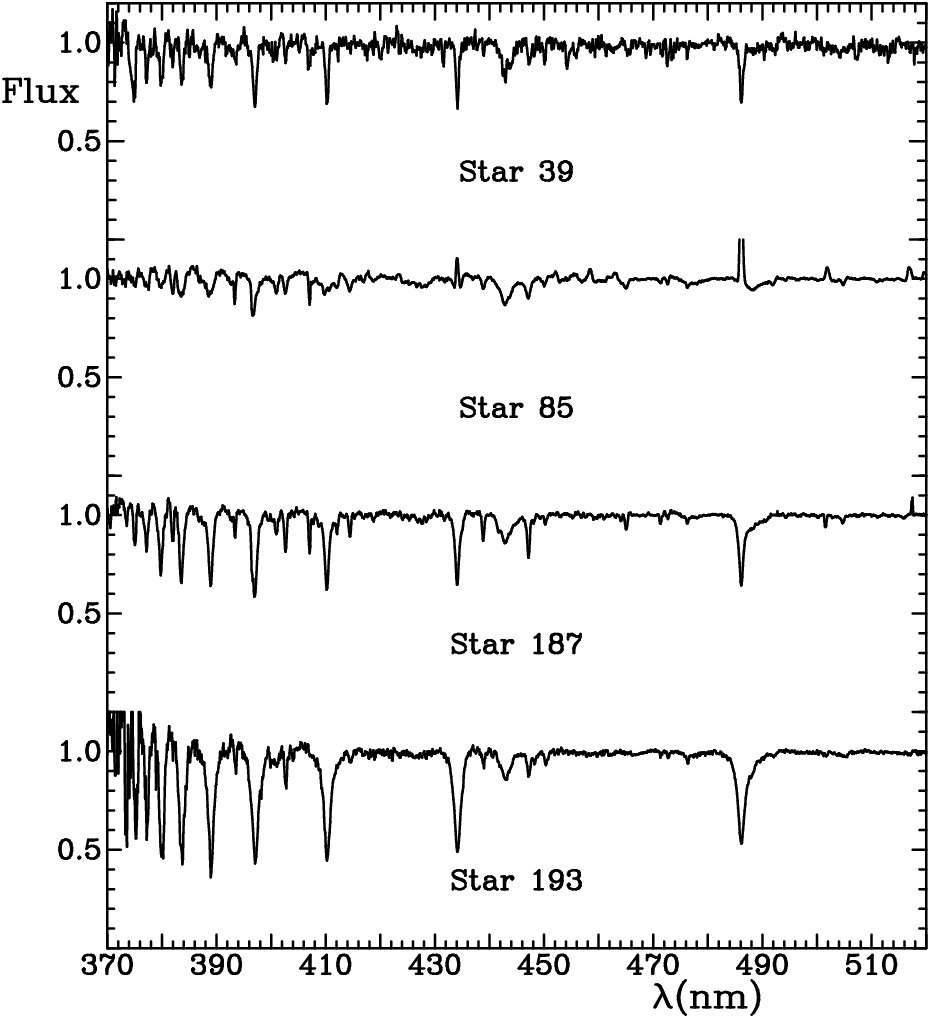}
\caption{Normalized spectra for bright stars in SAI~113.}
\label{fig4}
\end{figure}

Spectra were also obtained for four of the bright stars in SAI~113 using the Gemini Multi-Object Spectrograph (GMOS) in long-slit mode on the 8-m telescope of Gemini South, Chile. The observations were obtained in poor weather service observing periods of July 2016 with a 0.75 arcsecond slit width and grating B600, providing a typical resolving power of $\sim1200$. The normalized spectra for the observed stars represent the average of four integrations and are presented in Fig.~\ref{fig4}, with spectral classifications and derived parameters summarized in Table~\ref{tab3}. The spectra have been cleaned of a few obvious artefacts not originating in the stars, and also restricted to the blue-green spectral region normally used for classification on the MK system.

\begin{deluxetable}{@{\extracolsep{-3.0mm}}ccccccccc}
\tabletypesize{\small}
\tablewidth{0pt}
\tablenum{3}
\tablecaption{Spectroscopic Results for SAI~113 Stars
\label{tab3}}
\tablehead{
\colhead{Star} &\colhead{{\it V}} &\colhead{{\it B--V}} &\colhead{{\it U--B}} &\colhead{Sp.T.} &\colhead{E$_{B-V}$} &\colhead{E$_{U-B}$} &\colhead{{\it V--M$_V$}}  &\colhead{{\it V$_0$--M$_V$}} }
\startdata
34\tablenotemark{a} &11.88 &0.72 &--0.31 &O8~V(n) &1.03 &0.82 &16.98 &12.87 \\
39\tablenotemark{b} &12.02 &0.63 &--0.39 &O7~V((f))n &0.95 &0.77 &16.82 &13.02 \\
85\tablenotemark{c} &12.82 &0.83 &--0.40 &B1~Vnne &0.95 &0.62 &16.62 &12.84 \\
187\tablenotemark{d} &13.88 &0.74 &--0.23 &B1.5~Vn &0.99 &0.67 &17.33 &13.38 \\
193\tablenotemark{e} &13.94 &0.77 &--0.15 &B2.5~Vn &0.99 &0.64 &16.74 &12.77 \\
\enddata
\tablenotetext{a}{2MASS~J10224377--5930182}
\tablenotetext{b}{2MASS~J10224096--5930305}
\tablenotetext{c}{2MASS~J10224440--5929394, HEN~3-406}
\tablenotetext{d}{2MASS~J10223910--5931080}
\tablenotetext{e}{2MASS~J10224587--5932420}
\end{deluxetable}

A fifth star was added to Table~\ref{tab3}, namely star 34, where the classification is from the third release of the Galactic O-Star Spectroscopic Survey, GOSSS \citep{mz16}. That survey also includes star~39; the latter being classified there as O7~V((f))z, virtually identical to the present classification, the ``z'' denoting strong He~II~$\lambda 4686$ absorption. Star 85 was detected as an emission-line star previously by \citet{he76}, and has the designation HEN~3-406. Reddenings for the hotter Be stars are affected by excess continuum emission according to \citet{sr76}, amounting to systematic excesses of 0.15 in E$_{B-V}$ and --0.05 in E$_{U-B}$. The reddenings for star~85 in Table~\ref{tab3} have therefore been adjusted for that effect. The adopted absolute magnitudes used to estimate distance moduli have also been adjusted according to prior experience indicating that dwarfs classified as nn, n, or (n), as well as dwarf Be stars, appear to have luminosities comparable to subgiants (class IV) rather than dwarfs (class V).

\section{Analysis of the Photometric Data}
The photometric {\it UBV} data for stars lying within the boundaries of Fig.~\ref{fig1} were tested by various means, with simple dereddening providing important results. It was found, for example, that dereddening solutions for stars with $V \ge 19$ were generally unrealistic when compared with those for brighter stars, so the analysis was restricted to objects with $V < 19$.

\begin{figure}[h!]
\epsscale{1.00}
\plotone{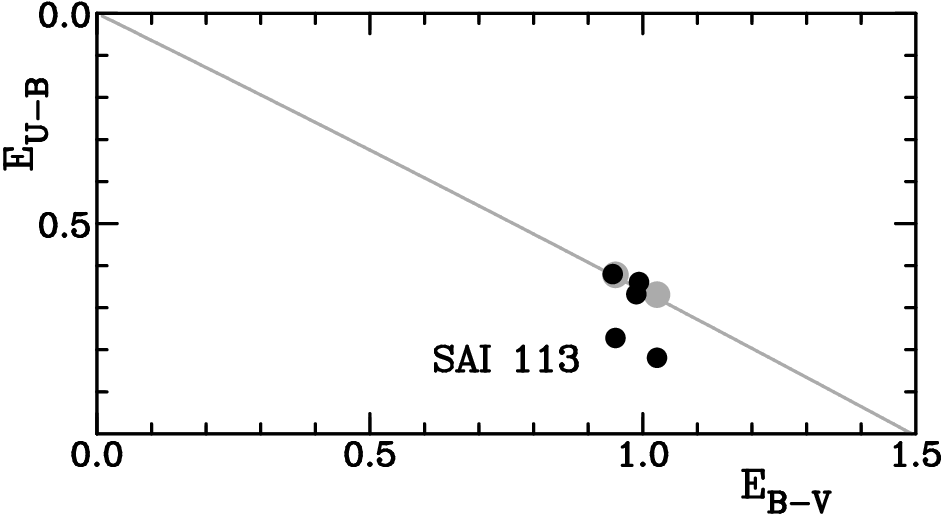}
\caption{{\it UBV} color excesses for spectroscopically observed stars in SAI~113 plotted with respect to a reddening line (in gray) described by E$_{U-B} = 0.64$ E$_{B-V} + 0.02$ E$_{B-V}^2$. Alternate E$_{U-B}$ reddenings for the O-stars adjusted by $-0.15$ are plotted as larger gray points.}
\label{fig5}
\end{figure}

The spectra for bright cluster stars listed in Table~\ref{tab3} have been classified on the MK system, here for 39, 85, 187, and 193, and by \citet{mz16} for 34, and, according to the discussion of \S2, generated the color excesses summarized in the table and plotted in Fig.~\ref{fig5}. The reddenings of the B-stars are a close fit to an extinction law described by: E$_{U-B} = 0.64$ E$_{B-V} + 0.02$ E$_{B-V}^2$, where the slope is that found previously for Carina clusters \citep{tu12a,ca13} and the curvature term is that obtained for Galactic fields by \citet{tu89}. The exceptions to such a close fit are the two O-type stars, which are offset by +0.15 in E$_{U-B}$ from the above reddening line. It may be that both are affected by continuum emission, although it can be noted that both lie in heavily-populated regions of SAI~113 and their photometry may simply be affected by crowding. It seems unlikely that the reddening law for O stars in the region differs from that for B stars. All five stars lie within $2^{\prime}$ of the center of SAI~113 and share common reddenings averaging E$_{B-V} = 0.98\pm0.03$ s.d. and common intrinsic distance moduli averaging $V_0$--$M_V = 13.03\pm0.25$, for $R_V = 4$.

Color-color diagrams also provide information about reddening laws, since reddened objects corresponding intrinsically to the hottest O-type stars, or to stars lying near the A3~V ``kink'' in the intrinsic color-color relation, provide direct information about the reddening line's likely slope E$_{U-B}$/E$_{B-V}$, which in turn appears to be related to the ratio of total-to-selective extinction $R_V = A_V$/E$_{B-V}$ \citep*{te14}.

\begin{figure}[th!]
\epsscale{1.00}
\plotone{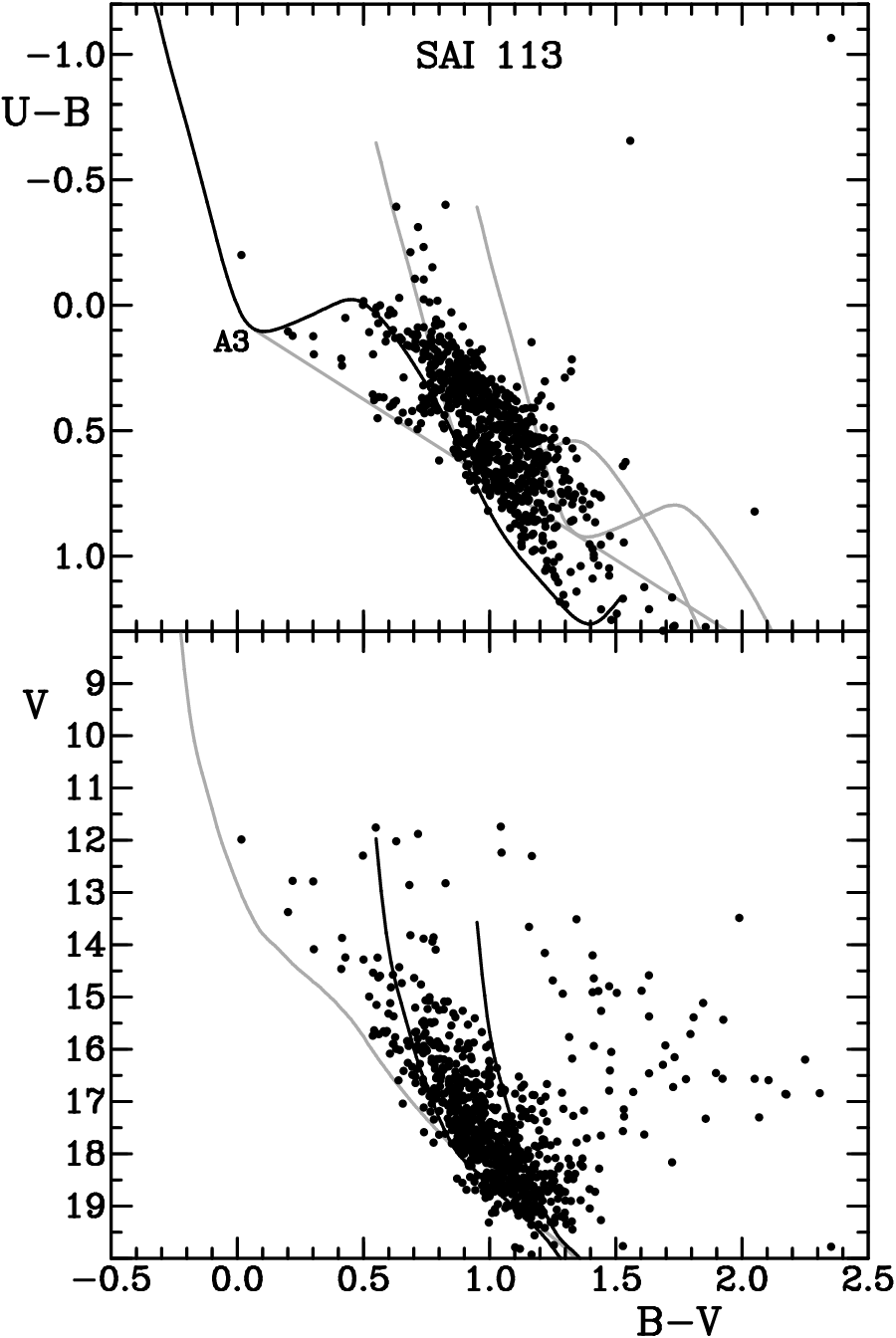}
\caption{{\it UBV} color-color (top) and color-magnitude (bottom) diagrams for stars within $6^{\prime}$ of the center of SAI~113. The black curve (upper) is the intrinsic relation for unreddened stars, while gray curves display its location for reddenings of E$_{B-V}=0.88$ (left) and $1.28$ (right). The straight line from the point denoted ``A3'' is a reddening line of slope E$_{U-B}$/E$_{B-V}=0.64$. The black curves in the lower diagram are ZAMS isochrones for the previous reddenings and $V_0-M_V=12.95$. The gray curve in the lower diagram is the ZAMS for E$_{B-V}=0.10$ and an intrinsic distance modulus of $V_0$--$M_V=11.83$.}
\label{fig6}
\end{figure}

The {\it UBV} data for stars lying within $6^{\prime}$ of the center of SAI~113 are analyzed in Fig.~\ref{fig6}, which plots color-color and color-magnitude diagrams from the CCD observations. It is evident from both diagrams that the cluster field appears to exhibit a small amount of differential reddening and extinction, with a minimum reddening of E$_{B-V} \simeq 0.88$ applying to reddened cluster B-type stars. Foreground and line-of-sight stars appear to exhibit reddenings of E$_{B-V} \le 0.20$, for the most part. In addition, the reddening for the main clump of B-type stars ranges from a minimum noted above to values apparently as large as E$_{B-V} \simeq 1.28$, although only late B-type stars exhibit reddenings that large; most are less reddened than E$_{B-V} \simeq 1.08$. The larger reddenings all apply to mid-to-late B-type stars, which typically exhibit large rotational velocities. They may exhibit an apparent excess reddening arising from rotational effects on the stellar continua, or possibly to a component of circumstellar reddening similar to that noted for late B-type stars in the open cluster Roslund~3 \citep{tu93,tu96}.

Note the clump of stars in the upper section of Fig.~\ref{fig6} that appear to be reddened from the A3 kink, matching a reddening law of slope E$_{U-B}$/E$_{B-V} = 0.64$, as found spectroscopically. An identical feature in the color-color diagrams for other stars in the field of Fig.~\ref{fig1} (see Figs.~\ref{fig8}~\&~\ref{fig9}) argues that the reddening law in the region matches that found in nearby Carina fields.

The zero-age main-sequence (ZAMS) relation tabulated by \citet{tu76a,tu79} is plotted in the lower portion of Fig.~\ref{fig6} for a reddening of E$_{B-V}=0.10$ and intrinsic distance modulus of $V_0$--$M_V=11.83$ (see later) in order to illustrate how less-reddened foreground stars are distributed in the color-magnitude diagram of SAI~113. Even with such stars removed from consideration, it is difficult to make a ZAMS fit for the B-type members of SAI~113 from the Fig.~\ref{fig6} data by themselves because of the deleterious effects of differential reddening and extinction. The plotted ZAMS fits for those stars are for $R_V = 4$ and $V_0$--$M_V=12.95$ at reddenings of E$_{B-V}=0.88$ and 1.28.

\begin{figure}[th!]
\epsscale{1.00}
\plotone{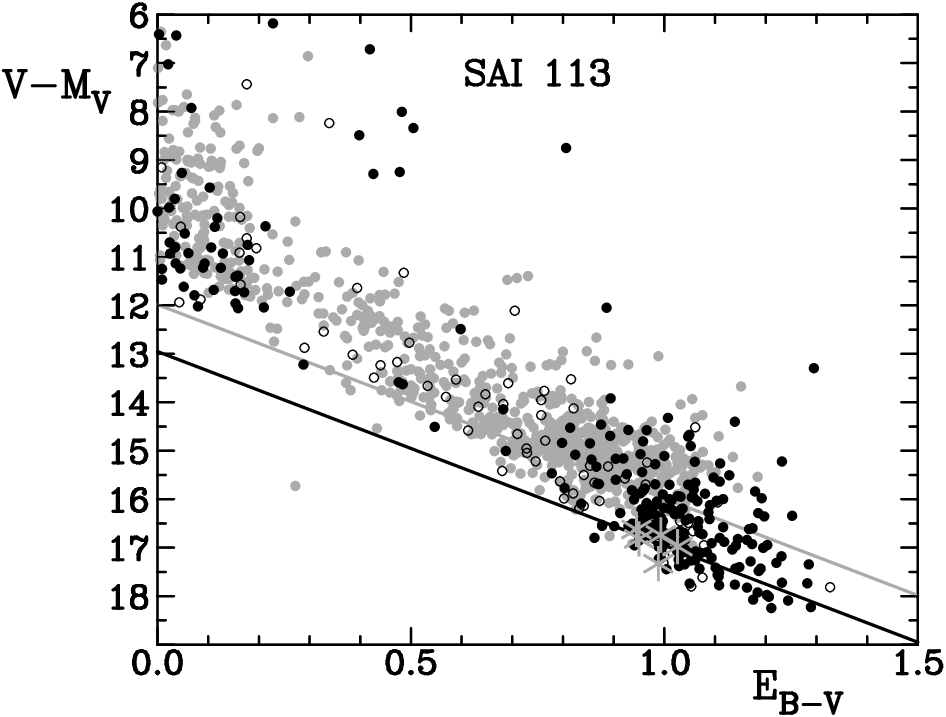}
\caption{A variable extinction diagram for SAI~113 tied to ZAMS values of $M_V$ for each star   (filled circles) and spectroscopic distance moduli (gray asterisks). Similar values are plotted for stars associated with van~Genderen~1 (open circles) and stars across the general field (gray points). The black relation of slope $R_V=$ E$_{B-V}/A_V=4$ and $V_0$--$M_V=12.95$ closely fits true ZAMS members of SAI~113, and the gray relation of identical slope with $V_0$--$M_V=11.99$ fits ZAMS members in van~Genderen~1 and the larger field.}
\label{fig7}
\end{figure}

\begin{figure}[th!]
\epsscale{1.00}
\plotone{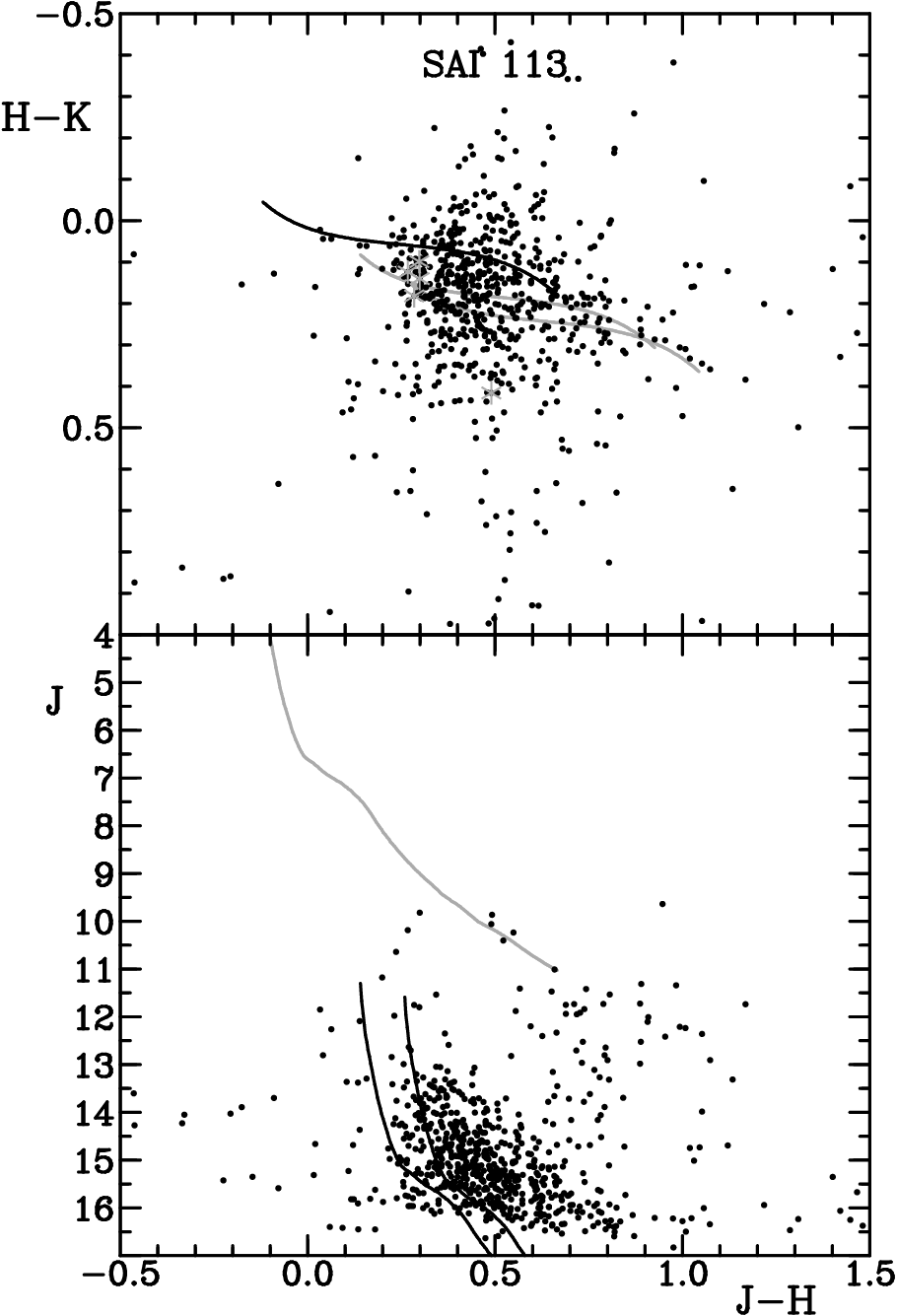}
\caption{2MASS {\it JHK}$_s$ color-color (top) and color-magnitude (bottom) diagrams for stars in SAI~113, with gray asterisks in the upper portion indicating spectroscopically observed stars. As in Fig.~\ref{fig6}, the intrinsic relation for unreddened stars is shown by the black curve (upper), and gray curves display its location for reddenings of E$_{B-V}=0.88$ and $1.28$. The black curves in the lower diagram are the ZAMS for the previous reddenings and $V_0-M_V=12.95$. The gray relation is the unreddened ZAMS for a distance of 10 pc.}
\label{fig8}
\end{figure}

\begin{figure}[th!]
\epsscale{1.00}
\plotone{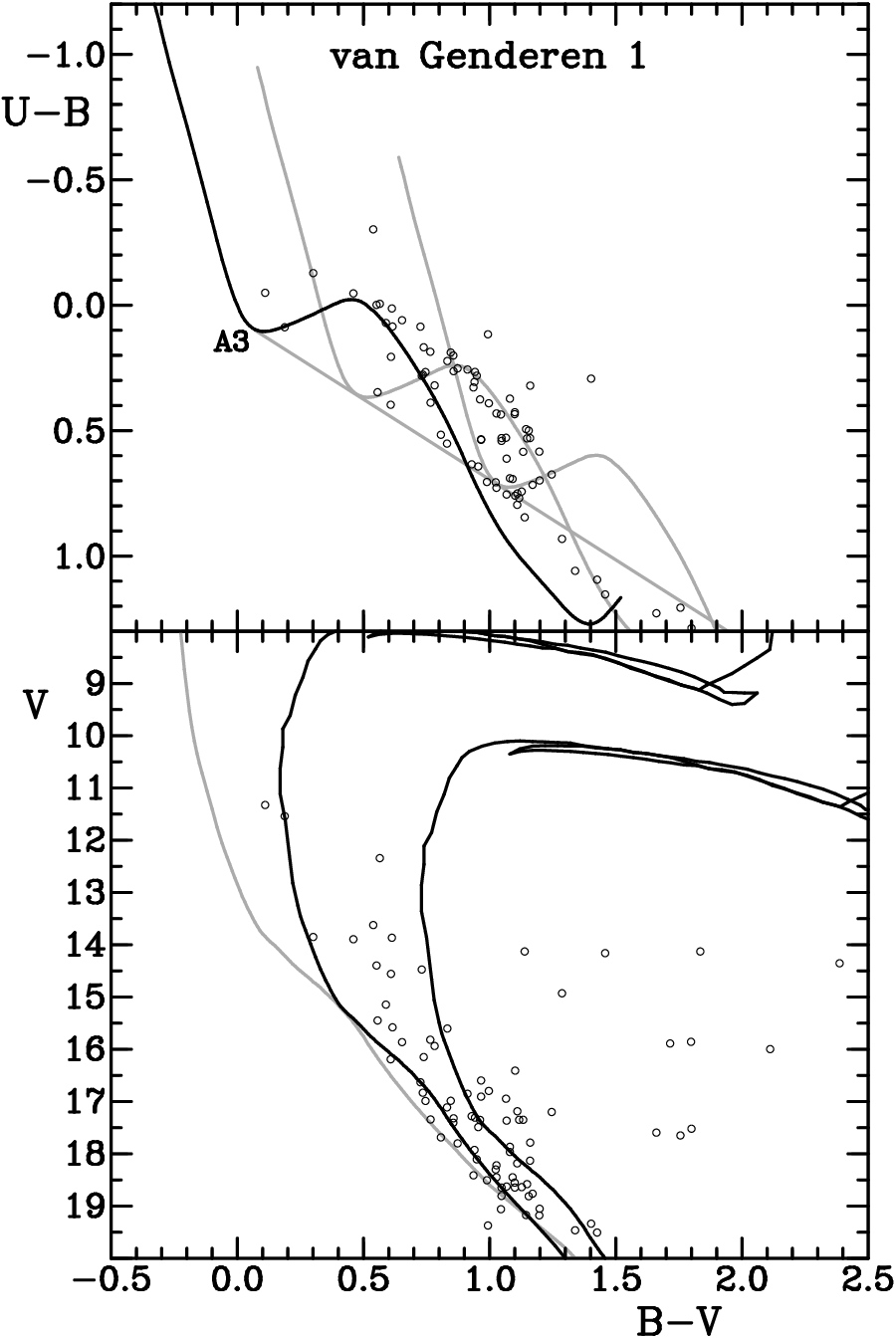}
\caption{{\it UBV} color-color (top) and color-magnitude (bottom) diagrams for stars in the core region of van~Genderen~1. Relations are as in Fig.~\ref{fig6}, except gray curves in the top portion display the intrinsic relation for reddenings of E$_{B-V}=0.41$ (left) and $0.97$ (right). The black curves in the lower diagram are $\log t= 7.1$ isochrones for the previous reddenings and $V_0-M_V=11.98$.}
\label{fig9}
\end{figure}

The implied presence of differential interstellar reddening in the field of SAI~113 from Fig.\ref{fig3} dictates that a detailed analysis of the photometric data requires use of the variable extinction method \citep[see, for example,][]{tu76a,tu76b}. By such a procedure the data for reddened early-type stars throughout the field of Fig.~\ref{fig1}, not just in SAI~113, were dereddened to the intrinsic zero-age zero-rotation main sequence \cite[ZAZRMS, see][]{tu16} using {\it UBV} reddening lines of slope E$_{U-B}$/E$_{B-V} = 0.64$. It was then possible to infer intrinsic {\it (B--V)}$_0$ colors and reddenings E$_{B-V}$, and corresponding apparent distance moduli $V$--$M_V$ from adoption of ZAMS absolute magnitudes for each star. All color excesses were adjusted to their equivalents for a B0 star using the relationship of \citet{fe63}. Where ambiguous dereddening solutions were evident, the likely ``best'' solution was guided by the 2MASS photometry for the star. Any ambiguities that remained were tested using alternate solutions, and in extreme cases were resolved by ignoring the affected star. Although some residual systematics may remain, they are considered minor. Results were isolated according to location in the field: main part of SAI~113, van~Genderen~1, and the remainder of the field lying outside the two main groupings. The combined data are shown in the variable-extinction diagram of the field plotted in Fig.~\ref{fig7}.

The value of $R_V$ for the field was established by identifying lower envelopes to the data for both stellar groups, with the dispersion in apparent distance moduli $V-M_V$ for each group gauged by eye to be less than $\pm0^{\rm m}.5$. For SAI~113 that corresponds to the lower section of filled circles in Fig.~\ref{fig7}, while for van~Genderen~1 it was necessary to identify a more luminous lower envelope to the open circles and gray filled circles. Both least squares and non-parametric techniques were applied to the 67 stars identified with the lower envelope of data for SAI~113, in order to reduce potential bias. Least squares methodology generally assumes no uncertainties in one co-ordinate, while non-parametric techniques make no assumptions about uncertainties in either co-ordinate. Averaging the results of both techniques weighted according to their resulting uncertainties yielded values of $R_V=4.19\pm0.22$ and $V_0$--$M_V=12.95\pm0.11$ (for $R_V=4.0 \pm0.1$) for SAI~113, corresponding to a distance of $3.90\pm0.19$ kpc. The data for 221 stars identified with the higher-lying lower envelope of data for van~Genderen~1 and the larger field yielded values of $R_V=4.06\pm0.05$ and $V_0$--$M_V=11.98\pm0.08$ (for $R_V=4.0 \pm0.1$) by the same methodology, corresponding to a distance of $2.49\pm0.09$ kpc.

The color difference method was also applied to the stars in Table~\ref{tab3}. Consistent estimates were obtained for stars 39, 187, and 193, yielding a value of $R_V=4.22\pm0.08$. The O8 dwarf, star 34, and Be star, 85, yielded larger values of $R_V$ near 5 and 6, which were considered to be affected by emission from circumstellar dust. A value of $R_V=4.0 \pm0.1$ was adopted for the entire region based upon the results of various estimates for {\it R} derived here and in other studies of stellar groups in the Carina field (see Table~\ref{tab4}).

Many of the same conclusions can be reached from comparable 2MASS color-color and color-magnitude diagrams for stars in SAI~113 plotted in Fig.~\ref{fig8}, where the intrinsic relations are from \citet{tu11}, converted from their equivalent reddenings in the {\it UBV} system. The addition of an unreddened ZAMS relation corresponding to a distance of 10 pc is used to indicate that a sizeable component of the color-magnitude diagram of Fig.~\ref{fig8} originates from unreddened late-type stars lying at distances ranging from 10--1000 pc along the line of sight to Carina. Otherwise, the conclusions reached from the {\it JHK}$_s$ data of Fig.~\ref{fig8} match those reached from the {\it UBV} data of Fig.~\ref{fig6}, although with some loss of clarity because of the larger photometric uncertainties, primarily in {\it H--K}.

\begin{deluxetable}{@{\extracolsep{-2.5mm}}lcl}
\tabletypesize{\small}
\tablewidth{0pt}
\tablenum{4}
\tablecaption{$R_V$ Estimates for Carina
\label{tab4}}
\tablehead{
\colhead{Method} &\colhead{$R_V$} &\colhead{Source} } 
\startdata
SAI~113~VE &$4.19\pm0.22$ &This paper \\
van~Gend~1~VE &$4.06\pm0.05$ &This paper \\
SAI~113~Col.~diffs. &$4.22\pm0.08$ &This paper \\
Ruprecht~91~VE &$3.82\pm0.13$ &\citet{te05} \\
Shorlin~1~CMDs &$4.03\pm0.08$ &\citet{tu12a} \\
IC~2581~VE &$3.77\pm0.19$ &\citet{ca13} \\
West~2~VE &$3.88\pm0.18$ &\citet{ca13} \\
West~2~Col.~diffs &$3.85\pm0.07$ &\citet{ca13} \\
West~2~Spec.~fits &$3.77\pm0.09$ &\citet{va13} \\
\\
Weighted Mean &$3.99\pm0.03$ & \\
\enddata
\tablecomments{VE = via variable-extinction analysis.}
\end{deluxetable}

A feature of Fig.~\ref{fig8} that matches that in Fig.~\ref{fig6} is the good fit of the least reddened cluster stars to the ZAMS. But the clump of more highly reddened cluster stars in Fig.~\ref{fig8} differs from that in Fig.~\ref{fig6} in containing many more stars redder than the ZAMS. The reason for that is not clear. Possibly it arises from systematic offsets in the 2MASS colors, or it may be related to the circumstellar reddening effect that appears in reddened B-type stars in Fig.~\ref{fig6}. It can be noted that the nuclear radius of $1^{\prime}.5$ for SAI~113 noted in \S~1 corresponds to only 1.7 pc at the distance of the cluster, somewhat smaller than values of 2--3 pc typical of most open clusters. SAI~113 is therefore likely to be a post core-collapse cluster, in other words highly evolved dynamically, consistent with the possible existence of circumstellar reddening for late B-type stars and the generally large main-sequence scatter in the cluster's unreddened color-magnitude diagram (Fig.~\ref{fig11}). 

\begin{figure}[th!]
\epsscale{1.00}
\plotone{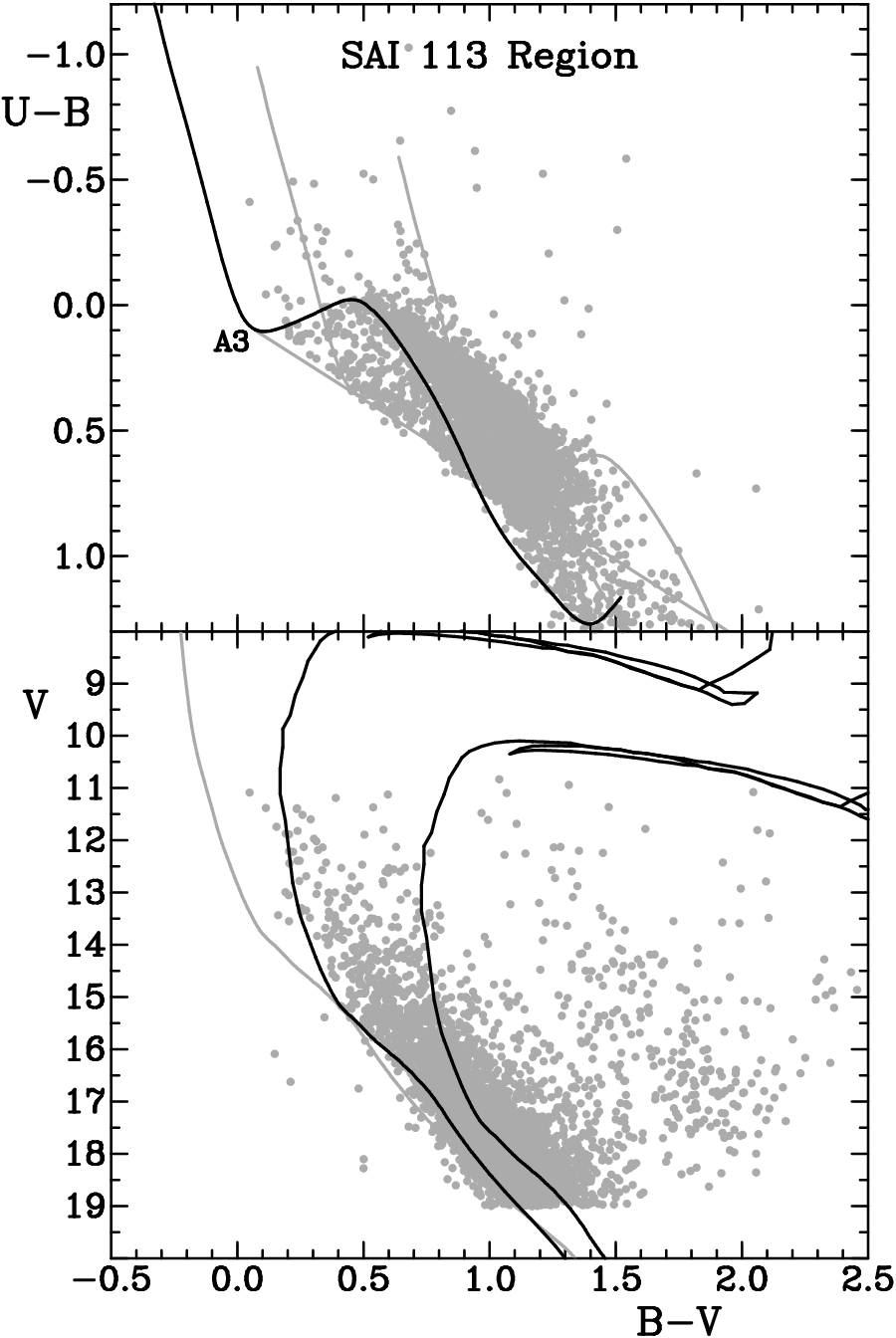}
\caption{{\it UBV} color-color (top) and color-magnitude (bottom) diagrams for stars in the general region surrounding van~Genderen~1 and SAI~113. Relations are as in Fig.~\ref{fig9}.}
\label{fig10}
\end{figure}

\begin{figure}[h!]
\epsscale{0.80}
\plotone{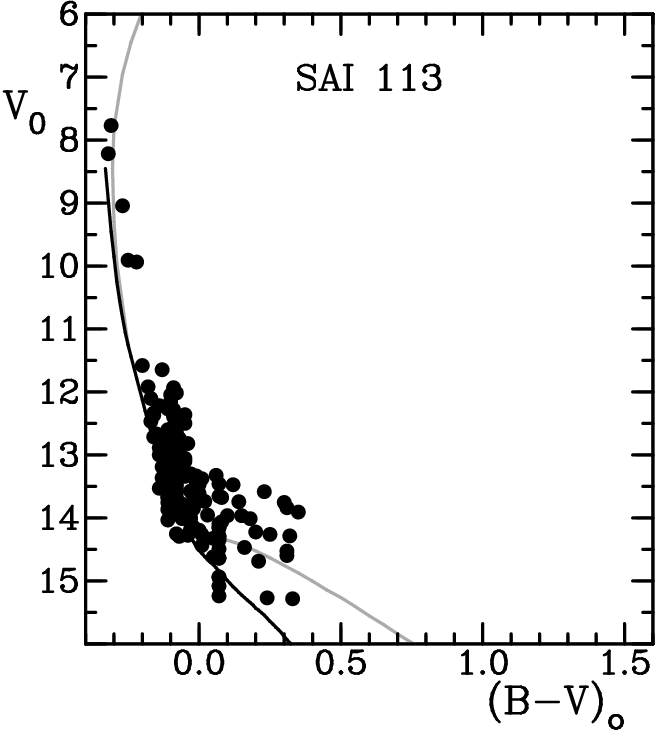}
\caption{Unreddened color-magnitude diagram for likely members of SAI~113 for an intrinsic distance modulus of $V_0$--$M_V=12.95$. The ZAMS is plotted as a black curve, an isochrone for $\log t = 6.75$ as a gray curve.}
\label{fig11}
\end{figure}

Similar plots of {\it UBV} color-color and color-magnitude diagrams are presented in Fig.~\ref{fig9} for the core region of van~Genderen~1, and Fig.~\ref{fig10} for brighter stars in the general field of Fig.~\ref{fig1}, namely the surroundings of van~Genderen~1 and SAI~113. Symbols in the two figures are identical to those used in Fig.~\ref{fig7}, in order to avoid ambiguities. Main sequence members of van~Genderen~1 are clearly less distant than those of SAI~113

As noted earlier, the least reddened stars in the variable-extinction diagram of Fig.~\ref{fig7} have intrinsic distance moduli of $V_0$--$M_V=11.83$, or 2.3 kpc. Yet there is a crude upper envelope to the ZAMS-fitted stars in Fig.~\ref{fig7} near $V_0$--$M_V\simeq10$, or 1 kpc, which is where the primary dust extinction begins, according to \citet{nk80}. Such complications are typical of variable-extinction analyses, and generally arise because of ambiguous dereddening solutions for some stars. The overall conclusions reached from Figs.~\ref{fig6},~\ref{fig7},~\ref{fig8},~\ref{fig9}, and \ref{fig10} are that differential reddening in SAI~113 is relatively small, amounting to only 0.40 in E$_{B-V}$, but that the foreground reddening in SAI~113 is larger than that in van~Genderen~1 or across much of the field. Such conclusions match those gleaned from a visual inspection of the field (Fig.~\ref{fig3}).

An extinction-free color-magnitude diagram is presented for likely members of SAI~113 in Fig.~\ref{fig11}. Members were selected using the results of the variable-extinction analysis (Fig.~\ref{fig7}) and spectral classification (Table~\ref{tab3}). Post-main-sequence and pre-main-sequence stellar evolutionary model isochrones were interpolated from the results of \citet{me93} and \citet{ps93}, as in \citet{tu12b}. The best match for the upper and lower ends of the main sequence was for $\log t = 6.75$, corresponding to an age of $5.6 \times 10^6$ yrs. The strongest restriction on that result is the presence of the O7 and O8 dwarfs (stars 34 and 39) in the cluster. The solid identification of faint cluster members is restricted by the fact that both foreground and background stars projected against SAI~113 share similar reddenings. Stars lying close to or slightly more luminous than the ZAMS in Fig.~\ref{fig11} are highly likely members. The advanced dynamical state of SAI~113 explains the large main-sequence scatter as the result of unresolved binaries or, more likely, rapidly rotating stars, possibly from close binary mergers \citep[see][]{tu96}.

\begin{figure}[th]
\epsscale{1.00}
\plotone{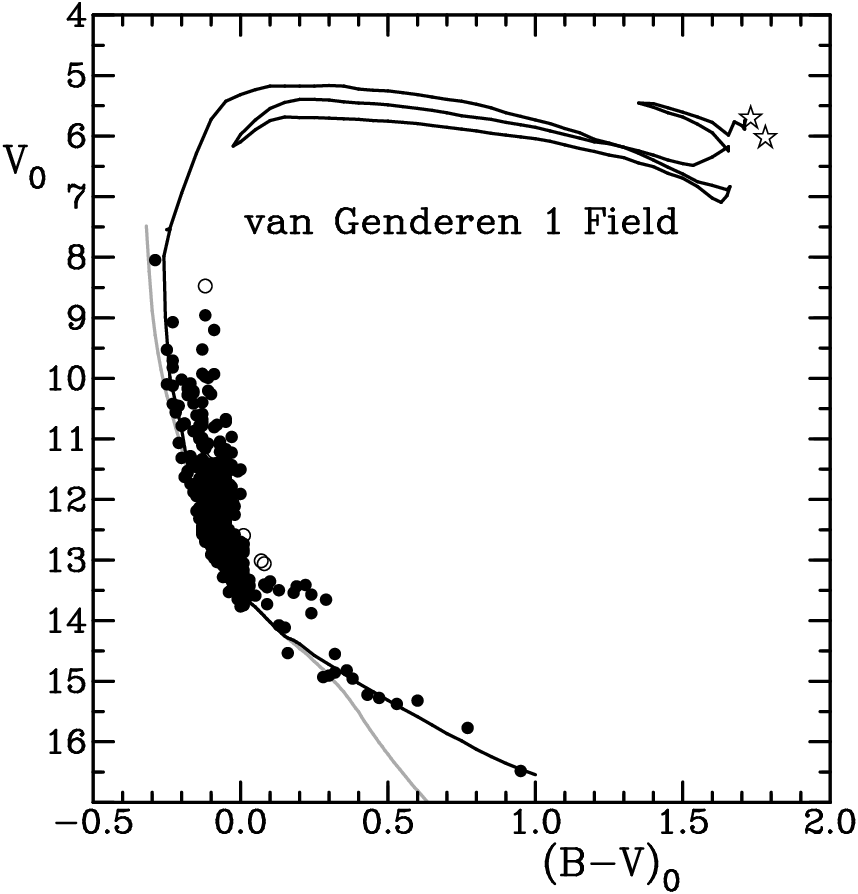}
\caption{Unreddened color-magnitude diagram for likely members of van~Genderen~1 lying in the field of SAI~113 for an intrinsic distance modulus of $V_0$--$M_V=11.98$. The ZAMS is plotted as a gray curve, an isochrone for $\log t = 7.1$ as a black curve, i.e. the inverse of Fig.~\ref{fig11}. Open circles represent foreground stars near the main section of SAI~113 that appear to lie foreground to the cluster, and star symbols represent the two M-supergiant variables lying south of van~Genderen~1.}
\label{fig12}
\end{figure}

An extinction-free color-magnitude diagram for likely outlying members of the group of stars associated with van~Genderen~1 is presented in Fig.~\ref{fig12}. The stars lie all across the field of Fig.~\ref{fig1}, and were selected based upon reddening and where they fell in the variable-extinction diagram (Fig.~\ref{fig7}). Once again, post-main-sequence and pre-main-sequence stellar evolutionary model isochrones were taken from those published by \citet{me93} and \citet{ps93}. The best match for the upper and lower ends of the main sequence was for $\log t = 7.1$, corresponding to an age of $1.3 \times 10^7$ yrs.

\begin{deluxetable}{@{\extracolsep{-2.5mm}}cccccccc}
\tabletypesize{\small}
\tablewidth{0pt}
\tablenum{5}
\tablecaption{Data for SRC Variables near van~Genderen~1
\label{tab5}}
\tablehead{
\colhead{Star} &\colhead{$\langle V \rangle$} &\colhead{$\langle B-V \rangle$} &\colhead{Sp.T.}  &\colhead{E$_{B-V}$} &\colhead{({\it B--V})$_0$} &\colhead{$V_0$} &\colhead{$M_V$} }
\startdata
CK~Car &7.87 &2.19 &M3.5~Iab &0.54 &1.73 &5.71 &--3.37 \\
EV~Car &7.87 &2.17 &M4.5~Iab &0.46 &1.78 &6.03 &--3.04 \\
\enddata
\end{deluxetable}

Data were included in Fig.~\ref{fig12} for the two M supergiant variables CK~Car and EV~Car, as summarized in Table~\ref{tab5}. The mean light data for the stars were taken from the Hipparcos catalogue \citep{esa97}, the spectral types are from the General Catalogue of Variable Stars \citep{sa12}, and intrinsic colors are those of \citet{le05}. The location of the two stars in the color-magnitude diagram of Fig.~\ref{fig12} agrees well with the $\log t = 7.1$ isochrone, strengthening the case that they may be outlying members of van~Genderen~1. Both stars are being investigated for periodicity by D.~G.~Turner and E.~Los using the Harvard College Observatory Photographic Plate Collection, with preliminary results indicating pulsation periods of 518 days for CK Car and 415 days for EV Car. If their outlying membership in van~Genderen~1 is confirmed, they become valuable calibrators for the period-luminosity relation for SRC variables (type C semi-regulars).

\section{Implications for Galactic Structure}
Although the derived parameters for SAI~113 and van~Genderen~1 are of interest, particularly with regard to their dynamical histories and stellar membership, of greater importance is how the discovery of a common value of $R_V=4$ throughout the Carina region (Table~\ref{tab4}) affects the delineation of Galactic spiral arm structure in the fourth quadrant.

Another concern is that the dust lanes running through Carina continue into the adjacent constellations, westward into Vela and Puppis, and eastward into Centaurus and Crux. That is the impression gained from visual inspection of blue light images of the Carina region and its environs. The results of \citet{nk80} suggest that the dust extinction in those regions increases noticeably at distances of about 1 kpc from the Sun, roughly midway between the Sun and the young star complexes in the same fields that help to delineate the Carina spiral arm. A similar impression is given by the more recent three-dimensional dust map of the Galaxy published by \citet{gr15}. Although the \citet{gr15} map does not include the fourth Galactic quadrant, high dust densities $\sim1$ kpc from the Sun along the borders of the adjoining first and third quadrants appear to be consistent with the \citet{nk80} map of nearby extinction, with a dust complex running diagonally across the region $\sim1$ kpc from the Sun. In such a situation one might expect $R_V \simeq 4$ throughout much of the fourth Galactic quadrant, at least for regions lying close to the Galactic equator where the dust clouds are concentrated visually.

It is of interest to note that the question of the appropriate value of $R_V$ in regions adjacent to Carina was addressed previously, at least for the Centaurus cluster Stock~16, by the study of \citet{va05}. They argue that $R_V \simeq 4$ at least in the regions surrounding the main cluster. That analysis, however, assumed a dependence of the color excess ratio E$_{V-I}$/E$_{B-V}$ on $R_V$ that is incorrect. As demonstrated by \citet{te11b}, the ratio E$_{V-I}$/E$_{B-V}$ is constant (value 1.26) for any value of E$_{U-B}$/E$_{B-V}$, the latter of which appears to be related to $R_V$ \citep{te14}. It is nevertheless possible to demonstrate that $R_V = 4$ in Stock~16, and that the same value applies in the Galactic plane over a large range of Galactic longitude.

\begin{figure}[bh]
\epsscale{1.00}
\plotone{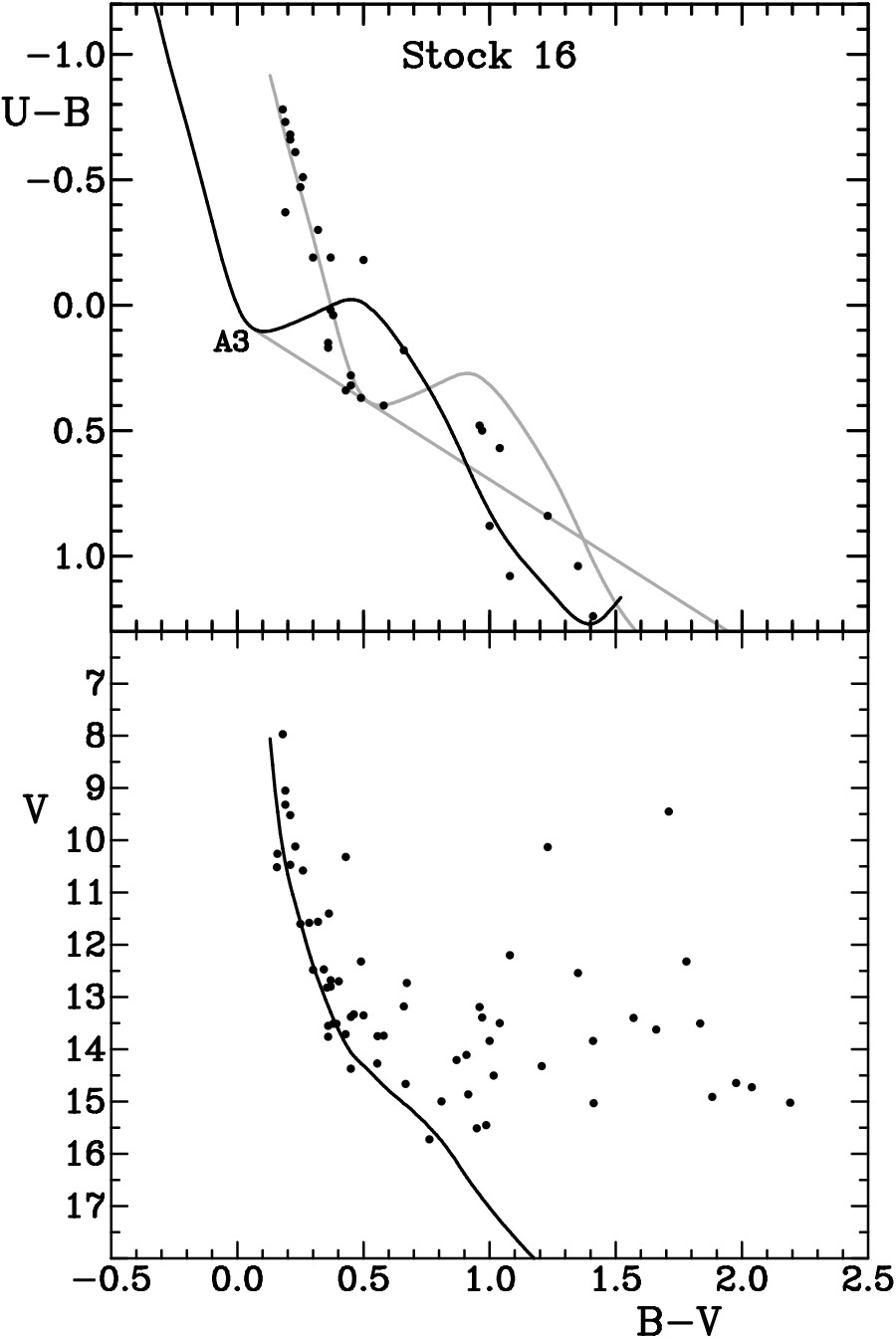}
\caption{{\it UBV} color-color and color magnitude diagrams for Stock~16, with relations as in Fig.~\ref{fig6}. The intrinsic relations in both cases are for a reddening of E$_{B-V} = 0.46$ and apparent distance modulus of $V$--$M_V = 12.56 \pm0.05$. The implied intrinsic distance modulus for $R_V = 4$ is $V_0$--$M_V = 10.72 \pm0.05$.}
\label{fig13}
\end{figure}

The Stock~16 case can be addressed in the same manner as our study of SAI~113, by means of {\it UBV} and {\it JHK}$_s$ photometry. Fig.~\ref{fig13} presents {\it UBV} data for Stock~16 stars, using photoelectric data from \citet{tu85} and CCD {\it BV} data from the APASS survey of the American Association of Variable Star Observers. A reddening law of slope E$_{U-B}/$E$_{B-V} = 0.64$ was adopted, consistent with the reddening of A3 dwarfs and the $R_V$ findings of \citet{va05}, when combined with the results of \citet{te14}. The best reddening for the B stars is E$_{B-V} = 0.46$ (top portion of figure), and the best ZAMS fit is for $V$--$M_V = 12.56 \pm0.05$. A similar result is obtained with the {\it UBV} data of \citet{va05}.

\begin{figure}[th]
\epsscale{1.00}
\plotone{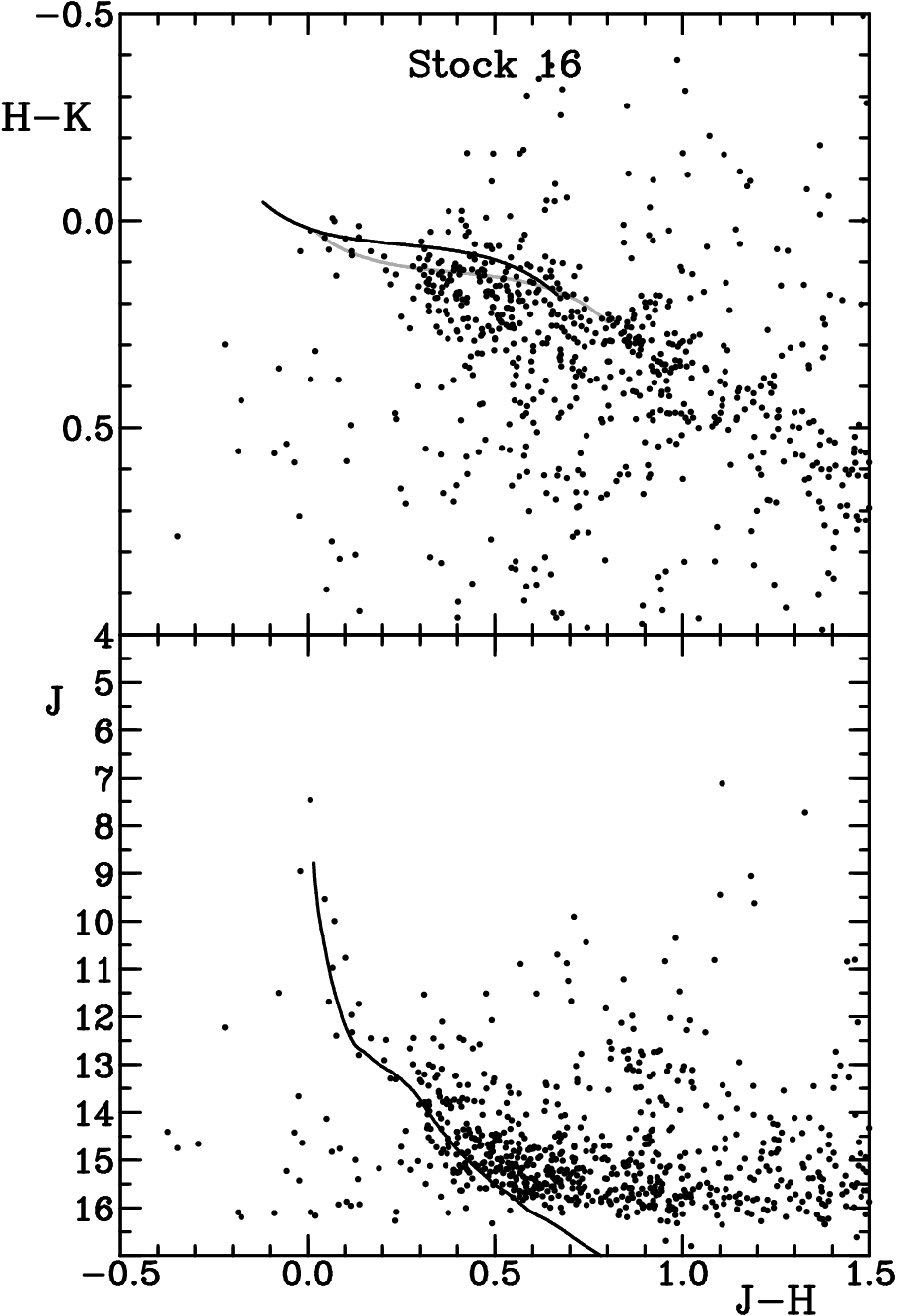}
\caption{{\it JHK}$_s$ color-color and color magnitude diagrams for Stock~16, with relations as in Fig.~\ref{fig8}. The intrinsic relations in both cases are for a reddening of E$_{J-H} = 0.136$ (E$_{B-V} = 0.46$) and apparent distance modulus of $J$--$M_J = 11.05 \pm0.05$, which corresponds to $J_0$--$M_J = 10.72 \pm0.05$, identical to the {\it UBV} results in Fig.~\ref{fig13}.}
\label{fig14}
\end{figure}

{\it JHK}$_s$ data for Stock~16 are shown in Fig.~\ref{fig14}. A reddening of E$_{B-V} = 0.46$ corresponds to  E$_{J-H} = 0.136$, which is a reasonably good fit to the color-color data for reddened B stars in the top portion of Fig.~\ref{fig14}. Large uncertainties in the 2MASS colors make it difficult to be more specific. The color-magnitude diagram for Stock~16 in the lower portion of Fig.~\ref{fig14} can be interpreted more readily. The best ZAMS fit here is for $J$--$M_J = 11.05 \pm0.05$. The corresponding intrinsic distance moduli, $V_0$--$M_V$ (for $R_V = 4$) and $J_0$--$M_J$, are both $10.72 \pm0.05$, corresponding to a distance of $1.393 \pm0.032$ kpc. Similar results can be obtained using the \citet{va05} observations.

Confirmation for the $R_V$ value is marginally possible using the variable-extinction method applied to the {\it UBV} data for ZAMS stars in Stock~16. The range of color excesses is small, $\sim0.12$ or slightly larger in E$_{B-V}$, but the methodology of Fig.~\ref{fig7} is still suitable. A similar analysis to that of Fig.~\ref{fig7} yielded a value of $R_V=4.04\pm0.49$ for Stock~16, matching the result obtained from ZAMS fitting in {\it JHK}$_s$ and {\it UBV}.

Although 2MASS observations are much less precise than typical photoelectric or CCD {\it UBV} photometry, they remain extremely useful for deriving reddenings and distances for open clusters \citep*{tu11,te11a,ma08,ma11,ma12}. Infrared photometry also enjoys the advantage of being less susceptible to variations in the extinction law, e.g. color excess ratios and $R_V$, that are so important for the analysis of optical photometry.

\begin{deluxetable}{@{\extracolsep{-1.5mm}}lrrcc}
\tabletypesize{\small}
\tablewidth{0pt}
\tablenum{6}
\tablecaption{New Distances to Southern Young Open Clusters
\label{tab6}}
\tablehead{\colhead{Cluster} &\colhead{$\ell$ (\degr)} &\colhead{$b$ (\degr)} &\colhead{$d_{\rm Lit}$(kpc)} &\colhead{$d_{\rm new}$(kpc)} }
\startdata
NGC~6611 &16.9540 &0.7934 &2.04 &\nodata \\
Waterloo~3 &242.562 &1.4434 &5.20 &2.63 \\
AQ~Pup &246.1562 &0.1061 &3.21 &\nodata \\
Pismis~11 &271.6567 &--0.7077 &3.60 &2.46 \\
Ruprecht~79 &277.0986 &--0.8180 &1.98 &3.09 \\
NGC~3105 &279.9147 &0.2636 &8.53 &2.70 \\
ASCC~58 &281.7116 &1.3281 &0.60 &0.67 \\
Loden~27 &282.1710 &--0.3346 &2.63 &2.20 \\
Loden~28 &282.2139 &--2.2145 &3.95 &2.34 \\
IC~2581 &284.5880 &0.0350 &2.87 &2.08 \\
Loden~112 &284.6870 &1.1641 &2.50 &2.13 \\
SAI~113 &285.0554 &--1.8883 &\nodata &3.90 \\
van~Genderen~1 &285.1145 &--2.0270 &\nodata &2.49 \\
Ruprecht~90 &285.4175 &--0.4372 &6.00 &5.14 \\
NGC~3293 &285.8527 &0.0717 &2.50 &2.25 \\
NGC~3324 &286.2283 &--0.1884 &3.12 &2.08 \\
Hogg~7 &286.3948 &--2.5335 &4.58 &NC \\
Trumpler~14 &287.4081 &--0.5774 &3.10 &2.17 \\
Trumpler~15 &287.4100 &--0.3688 &2.60 &2.06 \\
Collinder~232 &287.4929 &--0.5436 &3.00 &0.99 \\
Trumpler~16 &287.6238 &--0.6517 &3.39 &2.06 \\
Collinder~228 &287.6677 &--1.0471 &2.51 &1.95 \\
Pismis~17 &289.4693 &0.1394 &3.50 &1.87 \\
Sher~1 &289.6373 &--0.2419 &5.88 &3.74 \\
Feinstein~1 &290.0251 &0.3847 &1.16 &2.10 \\
Loden~306 &290.5547 &--0.7845 &2.00 &1.83 \\
Shorlin~1 &290.6314 &--0.9072 &\nodata &2.94 \\
NGC~3572 &290.7047 &0.2022 &2.00 &1.83 \\
NGC~3766 &294.1169 &--0.0304 &2.21 &1.77 \\
NGC~4103 &297.5736 &1.1551 &1.63 &1.52 \\
Hogg~15 &302.0474 &--0.2417 &3.20 &1.96 \\
NGC~4755 &303.2057 &2.5089 &1.98 &1.80 \\
Stock~16 &306.1486 &0.0631 &1.90 &1.39 \\
Hogg~16 &307.4772 &1.3372 &1.59 &1.75 \\
Collinder~272 &307.5947 &1.2016 &2.10 &1.79 \\
NGC~6383 &355.6897 &0.0412 &0.99 &0.96 \\
\enddata
\tablecomments{NC = no change from literature value.}
\end{deluxetable}

As a test of the conclusions reached here, new distances were derived for young open clusters, mostly in the fourth Galactic quadrant, as a means of determining if the traditional picture of spiral arm structure in that quadrant is affected. They were added to similar estimates summarized in \citet{tu12a}. As in that study, the distances were mainly tied to 2MASS {\it JHK}$_s$ photometry, but also including existing {\it UBV} photometry in the literature	 and, where necessary, CCD {\it BV} observations from the APASS survey. The combined results are summarized in Table~\ref{tab6}, and include a few clusters in the Puppis region. The 2MASS and APASS data for the cluster Hogg~7 were not deep enough to establish a unique solution, mainly because of large differential reddening in the field. The distance to the cluster was therefore left unchanged. 

\begin{figure}[bh]
\epsscale{1.00}
\plotone{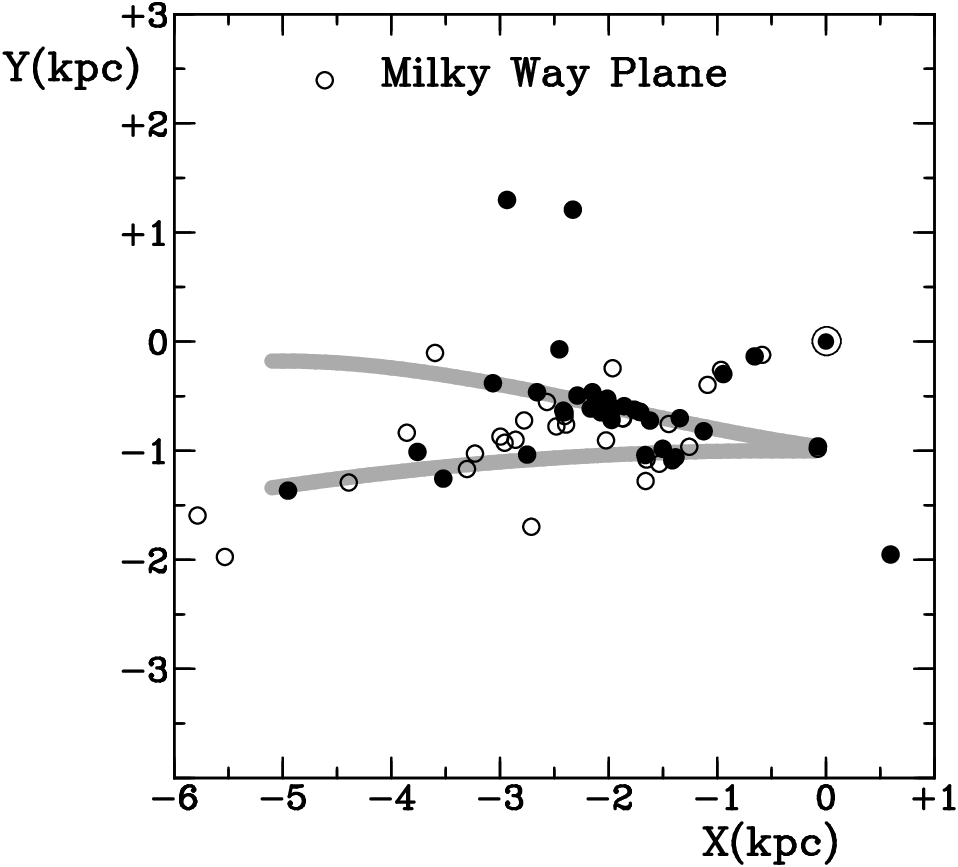}
\caption{A plot of Galactic plane projection for the young clusters summarized in Table~\ref{tab6}, with new distances denoted by filled circles, literature distances by open circles, and a sun-symbol indicating the location of the Sun. Gray curves are used to suggest how young open clusters may be connected to map spiral arms in the fourth quadrant.}
\label{fig15}
\end{figure}

The results are depicted in Fig.~\ref{fig15}, which displays the spatial location in the Galactic plane of Table~\ref{tab6} clusters, along with a schematic to illustrate the trends indicative of spiral arm features. In this picture the Carina arm does not run along the line of sight to the Carina clusters, but rather crosses it in near-linear fashion into Vela. Presumably the Carina arm connects to the Sagittarius arm, as generally  assumed, although in Fig.~\ref{fig15} there also appears to be a spur running from it inside the Carina arm. With the older distances to young clusters in the fourth quadrant, the picture of spiral arm structure is much different and less clear.

It may not be possible to confirm the picture of spiral arm structure in the fourth quadrant from 21-cm H~I data, given that previous studies have used erroneous corrections of the H~I radial velocities to the Local Standard of Rest \citep{tu14}. Additional tests are needed, however.

\section{Summary}
The present study was initiated in order to provide updated parameters for the young open cluster SAI~113 (Skiff~8) in Carina, such parameters having been left unresolved in the \citet{gl10} study. The field of the CCD {\it UBVI}$_C$ observations included the nearby group van~Genderen~1, which was examined as well. Both young groups are relatively young, with ages of $5.6 \times 10^6$ years and $1.3 \times 10^7$ years, respectively, lying at distances of $3.90 \pm0.19$ kpc and $2.49 \pm0.09$ kpc. SAI~113 displays evidence for substantial dynamical evolution, and appears to have begun dissolution into the general field. The group designated as van~Genderen~1 appears to be the remains of a former star cluster that has already lost most of its members, which are spread across the surrounding field and nearby SAI~113. Two potential outlying members of van~Genderen~1 are the M supergiant variables CK~Car and EV~Car, which could become valuable calibrators for the SRC period-luminosity relation.

An important consequence of the photometric study of the cluster fields is the confirmation that the extinction law is described by a reddening relation of slope E$_{U-B}/$E$_{B-V} = 0.64$ and a ratio of total-to-selective extinction of $R_V = 4$, identical to what has been found for other young clusters in Carina. Moreover, such ``anomalous'' extinction appears to extend to adjacent regions of the Galactic plane, affecting how we view the spiral arm structure in the fourth Galactic quadrant. It appears that the Carina spiral feature does not extend along the Galactic line of sight into Carina, but crosses it in roughly linear fashion, continuing into Vela. The arm is only about 2 kpc distant in this direction, rather than 2.5--3.0 kpc as sometimes believed \citep[see][]{dh97}, with the extinction apparently originating at distances of about 1 kpc.

Clearly, how we interpret extinction in various segments of the Galactic plane directly affects the accurate derivation of distances to stellar groups such as the young open clusters normally employed as spiral arm tracers. That point was first made by \citet{jo68} a half century ago in connection with the Galaxy's Perseus arm, but is still appropriate today, at least with regard to how spiral features are mapped in the fourth Galactic quadrant.

\subsection*{ACKNOWLEDGEMENTS}
This publication makes use of data products from the Two Micron All Sky Survey, which is a joint project of the University of Massachusetts and the Infrared Processing and Analysis Center/California Institute of Technology, funded by the National Aeronautics and Space Administration and the National Science Foundation. The study is also based partly on observations obtained at the Gemini South Observatory [program GS-2016A-Q-109, P.I. Alejo], which is operated by the Association of Universities for Research in Astronomy, Inc., under a cooperative agreement with the NSF on behalf of the Gemini partnership: the National Science Foundation (United States), the National Research Council (Canada), CONICYT (Chile), Ministerio de Ciencia, Tecnolog\'{i}a e Innovaci\'{o}n Product\'{i}va (Argentina), and Minist\'{e}rio da Ci\^{e}ência, Tecnologia e Inova\c{c}\~{a}o (Brazil).

\end{document}